\documentclass[aps,prd,reprint,superscriptaddress,preprintnumbers]{revtex4-2}
\pdfoutput=1
\usepackage{amsfonts, amssymb, amsmath}
\usepackage{mathrsfs}
\usepackage{hyperref}
\usepackage{tikz}
\usepackage[table]{}
\usepackage{empheq}
\usepackage[normalem]{ulem}

\usepackage{tensor}

\usepackage{color}
\definecolor{dark-gray}{gray}{0.20}
\definecolor{gray}{gray}{0.30}
\definecolor{light-gray}{gray}{0.80}
\definecolor{dark-red}{rgb}{0.7,0,0}
\definecolor{dark-green}{rgb}{0.1,0.4,0}
\definecolor{dark-blue}{rgb}{0.3,0.3,0.7}
\definecolor{light-blue}{rgb}{0.8,0.8,1}
\definecolor{blue}{rgb}{0,0,1}
\definecolor{red}{rgb}{1,0,0}
\definecolor{green}{rgb}{0,1,0}

\usepackage{hyperref}
\hypersetup{
    colorlinks=true,
    linkcolor=dark-blue,
    citecolor=dark-red,
    urlcolor=dark-blue,
    linktoc=page
}

\newcommand{\be}{\begin{equation}}
\newcommand{\ee}{\end{equation}}
\newcommand{\bea}{\begin{eqnarray}}
\newcommand{\eea}{\end{eqnarray}}

\renewcommand{\Im}{\text{Im}}

\newcommand\nn{\nonumber}

\usepackage{tikz}

\usepackage{xcolor}

\newcommand\fft[2]{\frac{#1}{#2}}

\begin{document}

\title{Microscopics of de Sitter Entropy from Precision Holography}

\date{\today}

\author{Nikolay Bobev}

\affiliation{Instituut voor Theoretische Fysica, KU Leuven, Celestijnenlaan 200D, B-3001 Leuven, Belgium}

\author{Thomas Hertog}

\affiliation{Instituut voor Theoretische Fysica, KU Leuven, Celestijnenlaan 200D, B-3001 Leuven, Belgium}

\author{Junho Hong}

\affiliation{Instituut voor Theoretische Fysica, KU Leuven, Celestijnenlaan 200D, B-3001 Leuven, Belgium}

\author{Joel Karlsson}

\affiliation{Instituut voor Theoretische Fysica, KU Leuven, Celestijnenlaan 200D, B-3001 Leuven, Belgium}

\author{Valentin Reys}

\affiliation{Instituut voor Theoretische Fysica, KU Leuven, Celestijnenlaan 200D, B-3001 Leuven, Belgium}
\affiliation{Universit\'e Paris-Saclay, CNRS, CEA, Institut de Physique Th\'eorique, 91191, Gif-sur-Yvette, France}

\begin{abstract}
\noindent We calculate quantum corrections to the entropy of four-dimensional de Sitter space induced by higher-derivative terms in the gravitational action and by one-loop effects. Employing the intertwinement in semiclassical gravity of Euclidean de Sitter and anti--de Sitter saddles, we embed effective de Sitter gravity theories in M-theory and express the entropy in terms of the regularized Euclidean anti--de Sitter action on an auxiliary $\mathrm{EAdS}_4 \times S^7/\mathbb{Z}_k$ background. We conjecture that the partition function of the holographically dual 3d ABJM CFT determines the explicit form of the corrections to the de Sitter entropy. This includes a logarithmic term, the coefficient of which, we show, agrees with an independent one-loop calculation around the $-S^4 \times S^7/\mathbb{Z}_k$ Euclidean de Sitter saddle. This provides evidence that the microscopic degrees of freedom behind the entropy of four-dimensional de Sitter space in gravitational theories with a holographic dual description are encapsulated by the path integral of the Euclidean CFT on the three-sphere.
\end{abstract}


\maketitle
\section{Introduction}\label{sec:intro} \noindent
Ever since the seminal work of Gibbons and Hawking \cite{Gibbons:1977mu,Gibbons:1976ue}, a microscopic understanding of de Sitter (dS) entropy has remained elusive. What are the quantum states or the degrees of freedom that the entropy supposedly counts, if it does count anything?

Gibbons and Hawking calculated the entropy $\mathcal{S}_{\mathrm{dS}}$ of dS space in semiclassical gravity. They put forward an expression for the quantum gravitational partition function $Z$ in terms of a Euclidean path integral and conjectured that \cite{Gibbons:1976ue}
\begin{equation}
    \label{SdS}
    \mathcal{S}_{\mathrm{dS}} = \log Z \,.
\end{equation}
Working in Einstein gravity with a positive cosmological constant, they found that the on-shell Euclidean action $I_{\mathrm{EdS}}$ of the round four-sphere saddle yields the familiar area law relating the entropy to one quarter of the dS horizon area in Planck units.

But what is this calculation telling us? The dS horizon is observer dependent, so it is not clear where the quantum microstates that the entropy might count could be located. Furthermore, the absence of a spatial boundary in dS space means that it is difficult to even define a thermodynamic ensemble. In fact, given that the Hamiltonian vanishes for cosmological spacetimes like dS space, the partition function simply reduces to the trace of the identity operator, leading some to wonder whether the entropy contains much physical information at all.

Here, we employ a chain of dualities to advance a microscopic interpretation of dS entropy, at least in some theories. Along the way, we show that the entropy encodes more information about the theory than what one might have thought at first sight.

We start by considering general four-derivative, purely gravitational theories in four dimensions with a positive cosmological constant. We view these, in an effective field theory spirit, as corrections to general relativity. Applying Wald's formalism~\cite{Wald:1993nt,Iyer:1994ys}, we calculate the corrections to the dS entropy induced by the higher-derivative (HD) terms in the action, namely a Weyl-squared, Gauss--Bonnet, and Ricci scalar squared term. The resulting expression for the entropy agrees with that obtained from a semiclassical evaluation of the Euclidean path integral on the four-sphere saddle in these HD theories.

Next, we take a closer look at the four-sphere saddle. It is well known that there is an intricate geometric connection in semiclassical gravity between Euclidean saddles that describe the birth of a dS universe in the Hartle--Hawking state and asymptotically Euclidean anti--de Sitter (EAdS) space \cite{Maldacena:2002vr,Harlow:2011ke,Hertog:2011ky,Hartle:2012qb}. In fact, as advocated in \cite{Hertog:2011ky,Anninos:2012ft,Hartle:2012tv}, this connection forms the basis of a particularly promising route toward a dS/CFT correspondence.
At the heart of it lies the observation that Euclidean dS (EdS) and EAdS can be thought of as two real sections of a single complex geometry. In effect, all saddle points defining the semiclassical Hartle--Hawking wave function admit a geometric representation in which their interior involves part of a deformed EAdS space, possibly with a complex matter profile \cite{Hertog:2011ky}. Working in this asymptotic EAdS representation of the wave function, the regularized asymptotic EAdS action of the saddles specifies the tree-level no-boundary measure.

\begin{figure}
    \centering
    \includegraphics[width=0.4\textwidth]{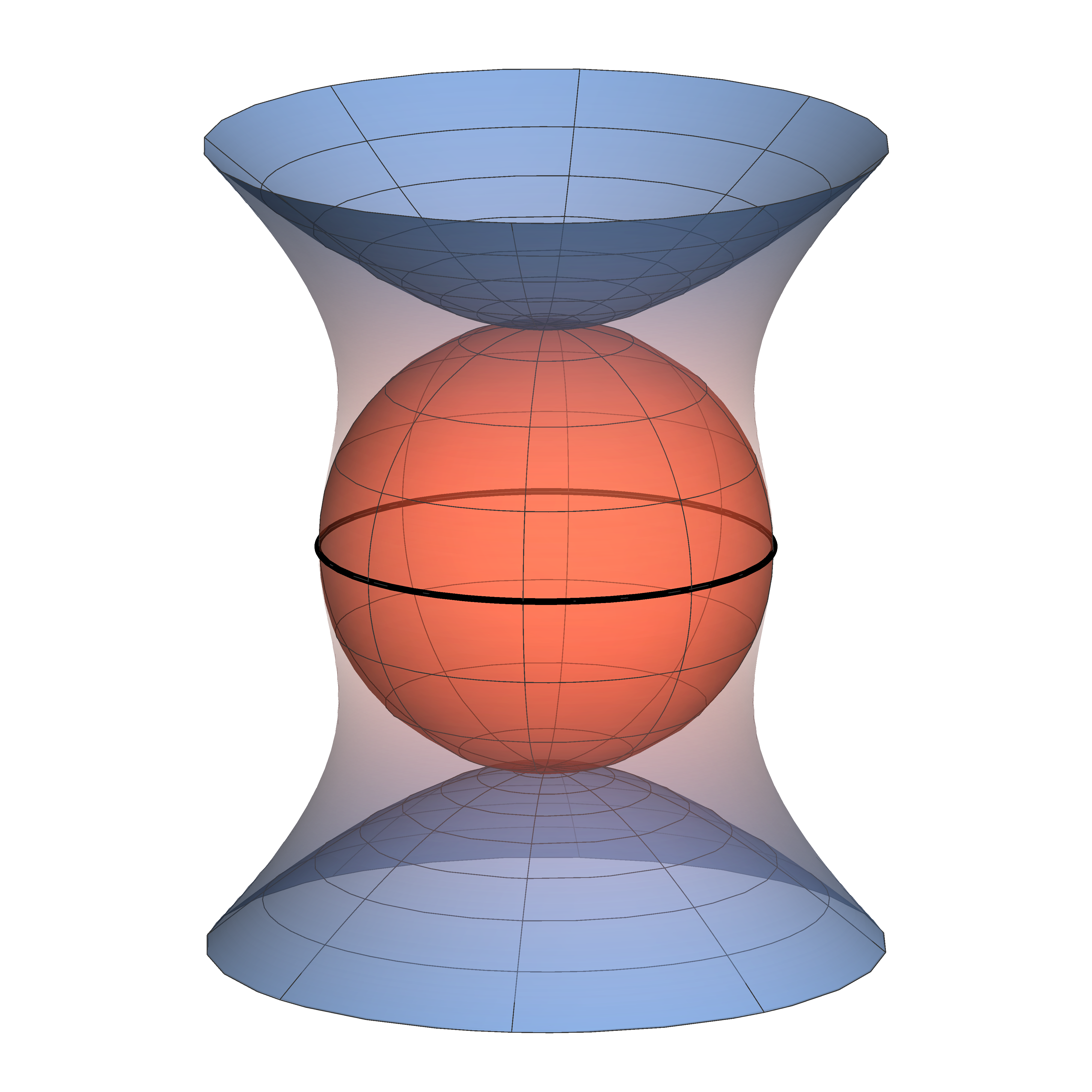}
    \caption{The round four-sphere saddle (red) can equally well be thought of as two copies of EAdS space (blue) emerging from both poles and connected through a complex transition region (gray).}\label{fig:S4EAdS4}
\end{figure}

The context we consider in this paper is similar in that the Euclidean form of the gravitational partition function in \eqref{SdS} is essentially the product of the complex conjugate branches of the Hartle--Hawking wave function, with no operators inserted anywhere. In particular, we can conceive of the four-sphere saddle in \eqref{SdS} as two copies of (part of) EAdS glued together through a complex transition region. Figure~\ref{fig:S4EAdS4} attempts to evoke these two alternative representations of the saddle. Even in the presence of HD terms in the action, we find that the corrected Euclidean action $I_{\mathrm{EdS}}$ of the round four-sphere can be obtained from the two regularized EAdS actions in the AdS representation of the saddle. Moreover, in the AdS domain, the HD gravity theory in dS we began with manifests itself as the bosonic action of four-dimensional gauged supergravity with four-derivative corrections. Indeed, the four-derivative Lagrangian for the propagating degrees of freedom of the $\mathcal N = 2$ gravity multiplet involves two extra real dimensionless constants, multiplying the Weyl-squared and Gauss--Bonnet terms, with the Ricci scalar squared term set to zero \cite{Bobev:2020egg,Bobev:2021oku}.

This brings us to the final step: Embedding this model in M-theory on $\mathrm{EAdS} \times S^7/\mathbb{Z}_k$ yields a third vantage point on the entropy in terms of the partition function of the holographic dual SCFT on the three-sphere. This third form, we conjecture, provides an exact quantum expression for the dS entropy. That is, we conjecture that the partition functions of dual CFTs on the three-sphere yield an exact expression for the dS entropy in certain four-dimensional gravitational theories.
To make this concrete and provide evidence for this conjecture, we use supersymmetric localization results in the 3d holographic SCFT to determine the two coefficients of the four-derivative terms in the supergravity action following~\cite{Bobev:2021oku}. In fact, working through the chain of dualities just described, the explicit form of the dual ABJM partition function also predicts the next-order (logarithmic) corrections to the dS entropy in these (embedded) HD theories, as well as all higher-order corrections.

Expressed in dS language, the embedding in M-theory on $\mathrm{EAdS} \times S^7/\mathbb{Z}_k$ elevates the four-sphere saddles to a $-S^4 \times S^7/\mathbb{Z}_k$ saddle in M-theory, where the minus sign refers to the overall signature with which the four-sphere enters. The upshot of our analysis, therefore, is that M-theory on backgrounds of this kind may well play a crucial role in cosmological applications of the theory.

Now, one might worry that the connection between EAdS and EdS space that we exploit holds on-shell only. One might suspect that this is a mere property of the tree-level path integral, albeit with HD corrections included, and dismiss this as a fancy way of repackaging an ``analytic continuation'' from EAdS to EdS. However, we show there is more to it. In the bulk, the logarithmic correction to the dS entropy arises from the one-loop determinants of kinetic operators of massless fields. Our conjectural relation between the $S^3$ partition function of the ABJM theory and the dS entropy yields a specific prediction for the coefficient of this logarithmic term. A gravitational calculation of this one-loop effect was performed in \cite{Bhattacharyya:2012ye} for  the $\mathrm{EAdS} \times S^7/\mathbb{Z}_k$ background and shown to agree with the field theory result. Along the same lines, we perform an independent calculation of the logarithmic correction to the dS entropy in M-theory on $-S^4 \times S^7/\mathbb{Z}_k$ and show that the result is twice that of the EAdS answer in perfect nontrivial agreement with our holographic conjecture for the dS entropy. We are therefore led to conclude that the intertwinement in quantum cosmology between EAdS and EdS space holds truth beyond tree level.

We continue our presentation in the next section with a discussion on how four-derivative corrections to general relativity modify the dS entropy. In Section~\ref{sec:contour}, we discuss an alternative description of the EdS saddle by exhibiting the saddle-point geometry along a complex time contour. Based on this analysis, we formulate our holographic conjecture for the exact dS entropy in terms of the $S^3$ partition function of a dual 3d SCFT in Section~\ref{sec:micro}. A nontrivial one-loop test of this conjecture is presented in Section~\ref{sec:1loop}. We conclude with some comments in Section~\ref{sec:discussion}. The appendix is devoted to some details on the calculation of regularized on-shell EAdS actions in the four-derivative gravity theory we study.

\section{Higher-derivative corrections to de Sitter entropy}\label{sec:HD} \noindent
We start by considering the most general four-derivative extension of general relativity in four dimensions, without matter but with a cosmological constant. The action $S_\pm$ reads, in Lorentzian signature,
\begin{widetext}
\begin{equation}
    \label{eq:Spm-4der-L}
    S_{\pm} = \int d^4x \sqrt{-g}\Bigl[\frac{1}{16\pi G_N}\Bigl(\pm R - \frac{6}{L^2}\Bigr)+(c_1 - c_2)\,C_{\mu\nu\rho\sigma}C^{\mu\nu\rho\sigma} + c_2\,\bigl(R_{\mu\nu\rho\sigma}R^{\mu\nu\rho\sigma} - 4R_{\mu\nu}R^{\mu\nu} + R^2\bigr) + c_3 R^2 \Bigr] \, .
\end{equation}
\end{widetext}
Here, $C_{\mu\nu\rho\sigma}$ is the Weyl tensor and $(c_1,c_2,c_3)$ are real constants that parametrize the three independent combinations of HD terms. At this stage, in an effective field theory spirit, we allow these to take arbitrary values, although they should of course be thought of as small compared to the dimensionless ratio $L^2/G_N$. That is, we view the above theory as a general gravitational correction to general relativity in the presence of a cosmological constant which we take to be positive, i.e.\ $\Lambda = 3/L^2 >0$.

The two actions $S_\pm$ in \eqref{eq:Spm-4der-L} describe the same theory expressed in different signatures, respectively mostly plus and mostly minus, i.e.\ $S_{+}[g_{\mu\nu}] = S_{-}[-g_{\mu\nu}]$. This is merely a convention, of course, and it may seem needlessly confusing to write down both choices. However, we shall be interested in saddle-point solutions of the semiclassical theory which tend to have Euclidean or even complex sections and hence intermingle various signatures. In the hope of clarifying what follows, we keep track of both signature conventions here.

The HD terms in the action \eqref{eq:Spm-4der-L} do not modify the Lorentzian dS solution of the two-derivative theory. In fact, neither the Gauss--Bonnet nor the Weyl-squared and $R^2$ terms induce corrections to any solution of the two-derivative equations of motion.
This being said, the HD terms do of course give rise to additional sets of solutions, as is well known e.g.\ from inflationary cosmology.

Working in global coordinates, the line element of the Lorentzian dS solution of the theories above is given by
\begin{equation}
    \label{eq:LdSpm-global}
    ds^2 = \pm\bigl[-dt^2 + L^2\cosh^2(t/L)\,d\Omega^2_{3}\,\bigr] \, ,
\end{equation}
where the overall sign corresponds to the sign choice in $S_{\pm}$.
In static-patch coordinates, we have
\begin{equation}
    \label{eq:LdSpm-static}
    ds^2 = \pm\Bigl[-\Bigl(1 - \frac{r^2}{L^2}\Bigr) dt^2 + \Bigl(1 - \frac{r^2}{L^2}\Bigr)^{-1} dr^2 + r^2d\Omega^2_{2}\,\Bigr] \, ,
\end{equation}
which makes apparent the presence of a cosmological horizon located at $r=L$.

The existence of a horizon suggests that one should associate an entropy to dS space \cite{Gibbons:1977mu}. At leading order in the derivative expansion, this entropy is simply given by the area of the horizon divided by $4G_N$. However, the HD couplings in \eqref{eq:Spm-4der-L} modify the entropy formula, even though they do not affect the dS solution itself.

In asymptotically Minkowski and AdS spacetimes with bifurcate horizons, Wald's formalism~\cite{Wald:1993nt,Iyer:1994ys} provides a natural way to account for HD corrections to the entropy associated with horizons. We lack a rigid derivation of this formalism in cosmological spacetimes~\cite{Note1}, but we nevertheless find it a useful starting point for our analysis.
In the context at hand, Wald's formalism holds that the dS entropy is given by the following integral:
\begin{equation}
    \label{eq:dS-Wald}
    \mathcal{S}_{\text{dS}} = \mp 2\pi\int_H d^2x\sqrt{\gamma}\,\frac{\delta \mathcal{L}_\pm}{\delta R_{\mu\nu\rho\sigma}}\,\epsilon_{\mu\nu}\epsilon_{\rho\sigma} \, ,
\end{equation}
where the integral is taken over the cosmological horizon $H$, $\gamma$ is the determinant of the metric induced on $H$, and $\epsilon_{\mu\nu}$ denotes the binormal to $H$ normalized such that $\epsilon_{\mu\nu}\epsilon^{\mu\nu} = -2$.
The overall sign on the right-hand side of~\eqref{eq:dS-Wald} reflects again the two signature choices we keep in our basket. Since physical quantities are obviously independent of the choice of signature, the entropy $\mathcal{S}_{\mathrm{dS}}$ on the left-hand side does not need a similar subscript. Using~\eqref{eq:Spm-4der-L}, a straightforward calculation gives
\begin{equation}
    \label{eq:ent-LdSpm-4der}
    \mathcal{S}_{\text{dS}} = \frac{\pi L^2}{G_N} + 64\pi^2 (c_2 +6 c_3) \, .
\end{equation}

We now compare this result with the entropy obtained from the on-shell action of the corresponding Euclidean solution.
The Euclidean counterparts of the Lorentzian actions in~\eqref{eq:Spm-4der-L} are given by
\begin{widetext}
\begin{equation}
    \label{eq:Spm-4der-E}
    S^E_{\pm} = \int d^4x \sqrt{g}\Bigl[-\frac{1}{16\pi G_N}\Bigl(\pm R - \frac{6}{L^2}\Bigr) - (c_1 - c_2)\,C_{\mu\nu\rho\sigma}C^{\mu\nu\rho\sigma} - c_2\,\bigl(R_{\mu\nu\rho\sigma}R^{\mu\nu\rho\sigma} - 4R_{\mu\nu}R^{\mu\nu} + R^2\bigr) - c_3 R^2 \Bigr] \, .
\end{equation}
\end{widetext}
Formally, these specify the quantum gravitational partition function $Z$ in terms of a Euclidean path integral \cite{Gibbons:1976ue}:
\begin{equation}
    \label{eq:Z-E}
    Z = \int \mathcal{D}g_{\mu\nu}\,e^{- S^E_{\pm}[g_{\mu\nu}]} \, .
\end{equation}
Of course, the very meaning of the path integral~\eqref{eq:Z-E} is obscure in a full quantum theory of gravity. It is one of our goals indeed to elucidate its meaning.

For now, we consider this quantity in a semiclassical approximation. The Einstein equations derived from~\eqref{eq:Spm-4der-E} admit a EdS solution with line element, in global coordinates, given by
\begin{equation}
    \label{eq:EdSpm-global}
    ds^2 = \pm\bigl[d\tau^2 + L^2\cos^2(\tau/L)\,d\Omega^2_{3}\,\bigr] \, ,
\end{equation}
where the coordinate $\tau$ lies in the range
\begin{equation}
    \label{eq:tau-segment}
    -\frac{\pi L}{2} \leq \tau \leq \frac{\pi L}{2} \, .
\end{equation}
We denote by $S^4$ the closed, conformally flat manifold specified by the metric with the upper sign in \eqref{eq:EdSpm-global},  and we denote by $-S^4$ its ``all-minus'' representation.
Note that the all-minus metric satisfies the criteria for a background in the gravity theory described by the action $S^E_-$ to be consistently coupled to a matter quantum field theory \cite{Kontsevich:2021dmb,Witten:2021nzp,Note6}, as the all-plus metric does for $S^E_+$.

Evaluating the actions~$S^E_{\pm}$ on their respective saddles \eqref{eq:EdSpm-global} yields
\begin{equation}
    \label{eq:Ipm-4der-E}
    I_{\text{EdS}} = -\frac{\pi L^2}{G_N} - 64\pi^2 (c_2 + 6c_3) \, ,
\end{equation}
where the second term represents the HD correction to the Euclidean action of EdS in general relativity.
Hence, in the saddle-point approximation, the entropy is given by
\begin{equation}
    \label{eq:Z-S}
    \mathcal{S}_{\text{dS}} = \log Z = \frac{\pi L^2}{G_N} + 64\pi^2 (c_2 + 6c_3)\, .
\end{equation}
Thus, we recover the Wald entropy~\eqref{eq:dS-Wald} in Lorentzian signature from a semiclassical evaluation of the Euclidean path integral on the four-sphere saddle. In the context of asymptotically Minkowski and AdS gravity, it is well-known that there is a relation between the entropy of gravitational horizons computed using the Wald formalism and the evaluation of a Euclidean on-shell action of the HD theory. The upshot of our short calculation above is that, although the Wald formalism may not enjoy a proper rigorous formulation in a cosmological setting, it can be applied to the dS horizon to yield results for the entropy consistent with the on-shell action of a general four-derivative gravitational theory with positive cosmological constant.

\section{A fresh look at the old four-sphere saddle}\label{sec:contour} 
\begin{figure}
    \centering
    \includegraphics{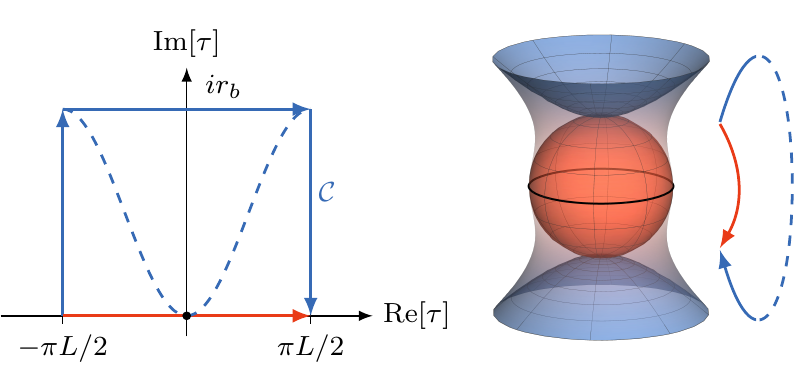}
    \caption{Left: two different integration contours, the segment in red and the blue solid contour $\mathcal{C}$, to evaluate the Euclidean on-shell action of the four-sphere saddle. Right: an illustration of the saddle in which the horizontal blue segment in the left panel is deformed to the dashed line. The black circle at the equator corresponds to the point at the origin in the left panel.}\label{fig:contour}
\end{figure}

\noindent
There is an alternative way to evaluate the Euclidean on-shell action of the four-sphere saddle that yields further insights. This involves deforming the contour for $\tau$ from the segment~\eqref{eq:tau-segment} of the real line, to the path $\mathcal{C}$ in the complex plane shown in Figure~\ref{fig:contour}.
Because the integrand in~\eqref{eq:Spm-4der-E} is analytic when evaluated on~\eqref{eq:EdSpm-global}, integrating along
$\mathcal{C}$ must yield the result~\eqref{eq:Ipm-4der-E}. However, the deformation exhibits an interesting geometric representation of the four-sphere saddle \cite{Hertog:2011ky,Harlow:2011ke}. Along the two vertical legs of $\mathcal{C}$, the metric \eqref{eq:EdSpm-global} is given by
\begin{equation}
    ds^2=\mp \left( dr^2+L^2\sinh^2(r/L)d\Omega_3^2\right)\, ,
    \label{EdS4:EAdS4}
\end{equation}
where $r = \Im(\tau)$ and the overall minus signature goes together with the overall plus signature in \eqref{eq:EdSpm-global} and vice versa. That is, along these contour legs, the EdS saddle behaves as EAdS.

Going around the entire contour $\mathcal{C}$ in Figure~\ref{fig:contour} yields a representation of the four-sphere saddle as two copies of (part of) EAdS up to a large radius $r_b$, glued together through a complex transition region, establishing a connection---on-shell for now---between EdS and EAdS. The reason that EAdS emerges in this context follows immediately from the fact that the action \eqref{eq:Spm-4der-E} of the EdS theory, in whatever signature, is closely related to the action of an EAdS theory in the opposite signature. Classically, the EdS and EAdS actions specify two distinct theories. Semiclassically, however, we see that they are connected and better viewed as two sections of one complexified theory.

We now evaluate the on-shell action \eqref{eq:Spm-4der-E} by integrating along $\mathcal{C}$. Along this contour, there is a natural split between the contributions from the vertical and horizontal parts of the contour:
\begin{equation}
    I_{\mathrm{EdS}} = I_\mathrm{L_{v}} + I_\mathrm{L_{h}} + I_\mathrm{R_{h}} + I_\mathrm{R_{v}} \, ,
\end{equation}
where the subscripts on the right-hand side label the four legs of the contour in what we hope is self-explanatory notation.
Evaluating the action, we find for the first vertical part
\begin{equation}
    \label{Lvert}
    I_\mathrm{L_{v}} = \Bigl(\frac{\pi L^2}{16G_N} + 4\pi^2 (c_2 + 6 c_3)\Bigr)\Bigl[9\cosh\Bigl(\frac{r_b}{L}\Bigr)-8 - \cosh\Bigl(\frac{3r_b}{L}\Bigr)\Bigr] \, ,
\end{equation}
and the first half of the horizontal leg gives
\begin{equation}
    \label{eq:mI-4der-Lhor}
    \begin{split}
        &I_\mathrm{L_{h}} = \Bigl(\frac{\pi L^2}{16G_N} + 4\pi^2 (c_2 + 6 c_3)\Bigr)\Bigl[\cosh\Bigl(\frac{3r_b}{L}\Bigr)-9\cosh\Bigl(\frac{r_b}{L}\Bigr)\Bigr] \\
        &\; -i\Bigl(\frac{\pi L^2}{16G_N} + 4\pi^2 (c_2 + 6 c_3)\Bigr)\Bigl[9\sinh\Bigl(\frac{r_b}{L}\Bigr) + \sinh\Bigl(\frac{3r_b}{L}\Bigr)\Bigr] \, .
    \end{split}
\end{equation}
Integrating over the second half of the contour produces the complex conjugate of the above expressions. Combining all contributions then yields
\begin{equation}
    \label{resultC}
    I_{\text{EdS}} = -\frac{\pi L^2}{G_N} - 64\pi^2 (c_2 + 6 c_3) \, ,
\end{equation}
which matches~\eqref{eq:Ipm-4der-E}, as it must.\\

The key point to glean from this alternative evaluation is that the imaginary part of the action integral along the horizontal leg does not contribute to the ``net result'' of the action $I_{\mathrm{EdS}}$, while the real part performs the holographic renormalization of the EAdS action resulting from the integral along the vertical legs. We demonstrate this explicitly for the Weyl-squared and Gauss--Bonnet terms in Appendix~\ref{App:regEAdS4}, where we show that the integration along the horizontal part of the contour produces what are known as the counterterms in the context of AdS/CFT and, importantly, nothing else. The large radius $r_b$ in Figure~\ref{fig:contour} acts as the radial cut-off of the EAdS region, and the horizontal leg of $\mathcal{C}$ regularizes the on-shell EAdS action from the vertical leg. In particular, expressed in the EAdS language of Appendix~\ref{App:regEAdS4}, the contributions \eqref{Lvert} and \eqref{eq:mI-4der-Lhor} can be written as
\begin{equation}
\begin{split}
    I_\mathrm{L_{v}} &= - I^{\text{reg}}_{\text{EAdS}} - I_\mathrm{ct}(r_b) + \mathcal{O}\left( e^{-r_b/L} \right)\,,\\
    I_\mathrm{L_{h}} &= + I_\mathrm{ct}(r_b) - I_\mathrm{ct}(r_b - i\pi L/2) + \mathcal{O}\left( e^{-r_b/L} \right)\,,
\end{split}
\end{equation}
where $I^{\text{reg}}_{\text{EAdS}}$ is the regularized on-shell action of the effective EAdS theory along the vertical leg of $\mathcal{C}$ induced by the theory \eqref{eq:Spm-4der-E} we started with and $I_\mathrm{ct}$ are the standard counterterms. Note that $I_\mathrm{ct}(r_b - i\pi L/2)$ is purely imaginary. The above relations are well-known from analogous dS/CFT studies in Einstein gravity (see e.g.~\cite{Hertog:2011ky}). What we have shown here is that they continue to hold in the presence of HD corrections.

Similar (complex conjugate) relations hold for the second half of the contour. The upshot, then, is that we can express the on-shell action \eqref{resultC} as
\begin{equation}
    \label{eq:mEdS-EAdS}
    I_{\text{EdS}} = -2\,I^{\text{reg}}_{\text{EAdS}} \, ,
\end{equation}
where the factor of 2 on the right-hand side arises from the fact that we have two copies of EAdS$_4$ along $\mathcal{C}$, glued together to yield a compact space with $S^4$ topology. This in turn yields yet another expression for the dS entropy:
\begin{equation}
    \label{eq:S-EAdS}
    \mathcal{S}_{\text{dS}} = 2\,I^{\text{reg}}_{\text{EAdS}} \, .
\end{equation}

This formula paves the way toward a concrete proposal for the microscopics behind the dS entropy using holography. We turn to this next.

\section{The microscopic nature of de Sitter entropy: a conjecture}\label{sec:micro} \noindent
To employ holography we first embed the effective gravitational theories we have so far considered in M-theory. This embedding allows us to think of the regularized EAdS on-shell action in~\eqref{eq:S-EAdS} as the saddle-point approximation of the eleven-dimensional, quantum gravitational path integral around $\mathrm{EAdS}_4 \times \text{SE}^7$, where the internal space is a seven-dimensional Sasaki--Einstein manifold.
To end up in the usual 11d conventions, in which both $\mathrm{EAdS}_4$ and the internal space have all-plus signatures, we continue with the lower sign from the previous two sections.

When the internal space is taken to be a particular smooth orbifold of the seven-sphere, holography predicts that the path-integral of M-theory around $\mathrm{EAdS}_4 \times S^7/\mathbb{Z}_k$ is given by the three-sphere partition function of the 3d $\mathcal{N}=6$ Chern--Simons-matter SCFT known as ABJM theory \cite{Aharony:2008ug}. This theory is a $\text{U}(N)_k \times \text{U}(N)_{-k}$ Chern--Simons-matter theory describing the low-energy limit of $N$ M2-branes probing a $\mathbb{C}^4/\mathbb{Z}_k$ singularity. The partition function of this theory on $S^3$ can be computed by supersymmetric localization in terms of matrix integrals \cite{Kapustin:2009kz}. In the large $N$ limit (with $k$ held fixed and finite), this partition function can be computed in terms of an Airy function plus nonperturbative corrections of the form $\mathcal{O}(e^{-\sqrt{N / k}})$ and $\mathcal{O}(e^{-\sqrt{N k}})$ \cite{Fuji:2011km,Marino:2011eh,Note2}:
\begin{equation}\label{eq:ABJMAiry}
\begin{split}
    &Z^{\text{ABJM}}_{S^3}(N,k) \\
    &= \Bigl(\frac{\pi^2 k}{2}\Bigr)^{\frac{1}{3}}\,e^{\mathcal{A}(k)}\,\text{Ai}\Bigl[\Bigl(\frac{\pi^2 k}{2}\Bigr)^{\frac{1}{3}}\Bigl(N - \frac{k}{24} - \frac{1}{3k}\Bigr)\Bigr] \\
    &\qquad + Z_{\text{n.p.}}(N,k) \, .
\end{split}
\end{equation}
The function $\mathcal{A}(k)$ is independent of $N$. Its explicit expression is available, see \cite{Hatsuda:2014vsa}, but we will not need it in what follows. The free energy $F^{\text{ABJM}}_{S^3} = -\log Z^{\text{ABJM}}_{S^3}$ can then be expanded at large $N$ and fixed $k$ using the known asymptotics of the Airy function. This yields
\begin{equation}
\begin{split}
    \label{eq:F-ABJM}
    F^{\text{ABJM}}_{S^3}(N,k)& = \frac{\pi\sqrt{2k}}{3}\,N^{\frac{3}{2}} - \frac{\pi(k^2 + 8)}{24\sqrt{2k}}\,N^{\frac{1}{2}}\\
    &\quad + \frac14\log N + \mathcal{O}(N^0) + \cdots \, ,
\end{split}
\end{equation}
where the terms on the right-hand side can be explicitly computed to arbitrarily high order in the $1/N$ expansion.\\

According to the AdS/CFT correspondence, this free energy maps to the action of a dual EAdS$_4$ saddle. The duality provides a dictionary that specifies bulk gravitational quantities in terms of the data $(N,k)$ that characterize the microscopic UV-complete dual theory. To illustrate this, let us recall how holography works in the large $N$ limit of the field theory, which corresponds to the leading two-derivative order in the bulk. In this limit, the length scale $L$ of the dual EAdS$_4$ is determined by the number of M2-branes in the 11d theory as (see e.g.~\cite{Marino:2011nm} for a review)
\begin{equation}
    \label{eq:L-N}
    (2\pi \ell_{\text{P}})^6 N = 6\,(2L)^6\,\text{vol}(X_7) \, ,
\end{equation}
where $\ell_{\text{P}}$ is the 11d Planck length and $\text{vol}(X_7)$ is the volume of the internal space $X_7 = S^7/\mathbb{Z}_k$. Similarly, the four-dimensional Newton constant is expressed in terms of $\ell_{\text{P}}$ by the standard Kaluza--Klein reduction of the two-derivative 11d action, which leads to
\begin{equation}
    \label{eq:holo-dict-leading}
    \frac{1}{G_N} = \fft{16\pi(2L)^7\text{vol}(X_7)}{(2\pi)^8\ell_P^9} =  \frac{2\pi^2\sqrt{6}}{9L^2\sqrt{\text{vol}(X_7)}}\,N^{\frac{3}{2}} \, .
\end{equation}
Using that vol$(S^7/\mathbb{Z}_k) = \tfrac{\pi^4}{3k}$, we thus find that the regularized two-derivative EAdS$_4$ on-shell action is
\begin{equation}
    \frac{\pi L^2}{2 G_N} = \frac{\pi\sqrt{2k}}{3}\,N^{\frac{3}{2}} \, ,
\end{equation}
which indeed matches the leading term in the free energy~\eqref{eq:F-ABJM}.

Away from the strict large $N$ limit, the holographic dictionary~\eqref{eq:holo-dict-leading} receives quantum corrections and the regularized on-shell action must now include contributions from the four-derivative couplings. Both of these effects combine at the subleading $\mathcal{O}(N^{1/2})$ order, and the results of~\cite{Bobev:2020egg,Bobev:2021oku} (see also Appendix~\ref{App:regEAdS4}) show that, remarkably, the holographic dictionary to this order reads
\begin{equation}\label{eq:holo-dict}
\begin{split}
    \frac{L^2}{G_N} + 64\pi c_2& = \frac{2\sqrt{2k}}{3}N^{\frac{3}{2}} - \frac{k^2 + 8}{12\sqrt{2k}}\,N^{\frac{1}{2}} + o(N^{\frac{1}{2}}) \, , \\
    c_3 &= 0\, .
\end{split}
\end{equation}
In fact, by considering the partition function of the ABJM theory on a more general squashed three-sphere and including mass deformations, it is possible to obtain the $1/N$ corrections to both $L^2/G_N$ and $c_2$ independently rather than the specific linear combination appearing above, see \cite{Bobev:2022eus,Note3}.
The results read, to all orders in the $1/N$ expansion,
\begin{equation}
    \label{eq:holo-dict-refined}
    \begin{split}
        \frac{L^2}{G_N} =&\; \frac{2\sqrt{2k}}{3}\Bigl(N - \frac{k}{24}\Bigr)^{\frac{3}{2}} \\
        =&\; \frac{2\sqrt{2k}}{3}\,N^{\frac{3}{2}} - \frac{k^2}{12\sqrt{2k}}\,N^{\frac{1}{2}} + \mathcal{O}(N^{-\frac{1}{2}}) \, , \\
        c_2 =&\; -\frac{1}{96\pi\sqrt{2k}}\Bigl(N - \frac{k}{24}\Bigr)^{\frac{1}{2}} \\
        =&\;  - \frac{1}{96\pi\sqrt{2k}}\,N^{\frac{1}{2}} + \mathcal{O}(N^{-\frac{1}{2}}) \, .
    \end{split}
\end{equation}
This is of course compatible with~\eqref{eq:holo-dict}. Note in particular that there are no $\log N$ terms in the expressions of the gravitational quantities $L^2/G_N$ and $c_2$ in terms of the microscopic data $(N,k)$. As explained in \cite{Bhattacharyya:2012ye}, the $\log N$ term in the free energy~\eqref{eq:F-ABJM} arises solely from a one-loop effect around the EAdS$_4$ saddle of the dual bulk theory, to which we return in Section~\ref{sec:1loop} below.\\

Turning now to the relation between the Wald entropy of dS space and the regularized on-shell action of EAdS that we derived in the previous section, the specific embedding of the effective HD theories in M-theory in combination with the corrected dictionary~\eqref{eq:holo-dict} allows us to write the dS entropy purely in terms of the ABJM data:
\begin{equation}
    \mathcal{S}_{\text{dS}} = \frac{2\pi\sqrt{2k}}{3}\,N^{\frac{3}{2}} - \frac{\pi(k^2 + 8)}{12\sqrt{2k}}\,N^{\frac{1}{2}} + o(N^{\frac{1}{2}}) \, .
\end{equation}
This is so far meant to be valid up to order $\mathcal{O}(N^{1/2})$ but it is tempting to conjecture that the correspondence holds to all orders in $1/N$ and thus that
\begin{equation}
    \label{eq:micro-dS}
    \mathcal{S}_{\text{dS}} = -2\log Z^{\text{ABJM}}_{S^3} \, .
\end{equation}
This conjecture, if true, provides a microscopic interpretation of the dS entropy in terms of the degrees of freedom of ABJM theory on the three-sphere.\\

Together with~\eqref{eq:F-ABJM}, our proposal~\eqref{eq:micro-dS} predicts a logarithmic correction to the dS entropy that reads
\begin{equation}
    \label{eq:log-test}
    \Delta\mathcal{S}_{\text{dS}} = \frac12\log N \, .
\end{equation}
In the next section, we show that this agrees with an independent one-loop calculation around the $-S^4 \times S^7/\mathbb{Z}_k$ background, which is the dS domain, along the real axis in Figure~\ref{fig:contour}, that appears in our embedding in 11d gravity. This constitutes a nontrivial test of the conjecture~\eqref{eq:micro-dS} to subsubleading order.

It is worth noting that logarithmic corrections to the entropy of dS space induced by loop effects in semiclassical gravity were recently studied in \cite{Anninos:2020hfj,Note4}.
The results in \cite{Anninos:2020hfj} for the coefficient of the log correction to the entropy differ from ours. This is not surprising. As emphasized by Sen in the context of black hole entropy, see for example \cite{Sen:2012dw}, the calculation of log corrections to the entropy is highly sensitive to the matter content of the effective semiclassical gravitational theory. Our setup has a natural embedding in a UV-complete theory of quantum gravity provided by M-theory and thus, as we discuss in the next section, the log correction to the dS entropy should be computed using the field content of the low-energy effective 11d supergravity theory. In contrast, the authors of \cite{Anninos:2020hfj} study the log corrections to the dS entropy when de Sitter space is viewed as a solution to 4d effective gravitational theory coupled to various matter fields.

\section{One-loop test} \label{sec:1loop} \noindent
On dimensional grounds, the logarithmic correction to the free energy of $-S^4$ arises from the one-loop determinants of kinetic operators for the massless fields. These operators can have zero modes which must be treated with care when computing such determinants. To do so, one usually splits the contributions of massless fields to the logarithmic correction into their nonzero mode and zero mode parts. An important simplification occurs when we expand the gravitational action to quadratic order around the $-S^4 \times S^7/\mathbb{Z}_k$ background of Euclidean 11d supergravity: it follows from the theory of heat kernels, see \cite{Vassilevich:2003xt} for a review, that the contribution from the nonzero modes vanishes, because the space is odd-dimensional. Therefore, we have to compute only the contribution from zero modes.

The massless fields of the 11d supergravity theory, viewed as a low-energy effective action of M-theory, are the metric $g_{MN}$, the gravitino $\psi_M$ and the three-form $C_{MNP}$. From the four-dimensional perspective, these fields can give rise to the 4d graviton $g_{\mu\nu}$ together with a collection of $p$-forms with $p=0,\ldots,3$ and spin $1/2$ and $3/2$ fermions. An important role is also played by the ghost fields required for quantization of the 11d physical fields, as we will see below.

It is perhaps useful to be more explicit about the 11d background that emerges in our setup. The 11d metric and 4-form on $-S^4 \times S^7/\mathbb{Z}_k$ are
\begin{equation}\label{eq:11dsol}
\begin{split}
    ds^2_{11} &= -\frac{1}{4}\left[d\tau^2+L^2\cos^2(\tau/L)d\Omega_3^2\right]\\
    &\quad+ L^2\left(d\psi+\sigma\right)^2+L^2ds^2_{\mathbb{CP}^3}\,,\\
    G_4 &= \fft{3}{8L}\text{vol}_4\,,
\end{split}
\end{equation}
where $\sigma$ is a 1-form potential for the K\"ahler form on the complex projective space $\mathbb{CP}^3$ with metric $ds^2_{\mathbb{CP}^3}$, $\text{vol}_4$ is the volume form for the metric in square brackets on the first line of \eqref{eq:11dsol}, and the $\mathbb{Z}_k$ orbifold acts on the angular coordinate $\psi$ which has period $2\pi/k$. One can check that this background solves the equations of motion of Euclidean 11d supergravity derived from the following action:
\begin{equation}
    S^E_{11\mathrm{d}}=-\frac{1}{16\pi G_N^{(11)}}\int\bigg[\mathord{*}R-\frac{1}{2}G_4 \wedge \mathord{*}G_4-\frac{i}{6}C_3\wedge G_4\wedge G_4\bigg]\,,\label{M:S-2der:E}
\end{equation}
where $G_4 = dC_3$, $*$ is the 11d Hodge star in Euclidean signature and $G_N^{(11)}$ is the 11d Newton constant. Importantly, this Euclidean action also admits another saddle given by the usual $\mathrm{EAdS}_4 \times S^7/\mathbb{Z}_k$ Freund--Rubin solution of 11d supergravity. These two different saddles can then be related by the 11d analog of the contour presented in Figure~\ref{fig:contour} which in turn provides a 11d realization of the discussion in Section~\ref{sec:contour}.

On the 11d space $-S^4 \times S^7/\mathbb{Z}_k$, the metric $g_{MN}$ does not have zero modes. This follows from the fact that there is no pure gauge mode with a non-normalizable gauge parameter on $-S^4 \times S^7/\mathbb{Z}_k$ because the space is compact. Quantization requires the introduction of a pair of vector ghost and anti-ghost associated with diffeomorphisms, and no residual gauge invariance is left, which implies that there are no ghost-for-ghosts (see e.g.~\cite{Townsend:1976vc}). The vector ghosts cannot have zero modes on spheres since the first Betti number vanishes, $b_1(S^{d>1}) = 0$. Similarly, there are no gravitino pure gauge modes $\psi_M \sim \mathcal{D}_M \epsilon$ with a non-normalizable spinor $\epsilon$ due to compactness, which means that the gravitino has no zero modes. Quantization requires the introduction of a pair of spinor ghost and anti-ghost, together with an additional spin $1/2$ ghost sometimes called the Kallosh--Nielsen ghost~\cite{Kallosh:1978de,Nielsen:1978mp}.
Regular solutions of the Dirac equation $\gamma^\mu\nabla_\mu \psi = i\lambda\psi$ on spheres exist only for $|\lambda| > 0$~\cite{Camporesi:1995fb}, and therefore the ghost fields associated to the gravitino do not contribute to the logarithmic correction to the free energy. Thus, the only contribution comes from the 3-form $C_{MNP}$.

In general, the logarithmic correction to the free energy due to a $p$-form on a $D$-dimensional background characterized by a length scale $\ell$ is given by~\cite{Bhattacharyya:2012ye}
\begin{equation}
    \label{eq:gen}
    \Delta F = \sum_j(-1)^j\bigl(\beta_{p-j} - j - 1\bigr)\,n_{\Delta_{p-j}}^0\,\log \fft{\ell}{\ell_{\text{P}}} \, ,
\end{equation}
where $\Delta_{p-j}$ is the Hodge--Laplace kinetic operator for a $(p-j)$-form, $n^0_{\Delta_{p-j}}$ is the associated number of zero modes, and $\beta_{p-j}$ is given by (see e.g.~\cite{Bhattacharyya:2012ye})
\begin{equation}
    \beta_s = \frac{D-2s}{2} \, .
\end{equation}
The above formula takes into account the ghosts required for the quantization of $p$-forms. In our case, $D=11$ and the background is $-S^4 \times S^7/\mathbb{Z}_k$ with a length scale $\ell = L$ common to the two factors. The zero modes of the Hodge--Laplace operator in the external space correspond to harmonic forms on $-S^4$. Thus, their number $n^0$ is nonvanishing only for 0-forms and 4-forms, since all Betti numbers vanish on the four-sphere except for $b_0 = b_4 = 1$. There are no 4-forms in the 11d field content discussed above. The 0-forms arise from the quantization of the 11d 3-form $C_{MNP}$, which produces two pairs of scalar ghost and anti-ghost fields~\cite{Siegel:1980jj}. This shows that, in the general formula~\eqref{eq:gen}, the nontrivial contribution comes from $p=3$ and $j=3$ with $n^0_{\Delta_0} = 2$. Putting things together,
\begin{equation}
    \Delta F_{\text{EdS}} = -\Bigl(\frac{11}{2} - 3 - 1\Bigr)\times 2\times \log \frac{L}{\ell_{\text{P}}} = -3 \log \frac{L}{\ell_{\text{P}}} \, .
\end{equation}
Lastly, we use the holographic dictionary~\eqref{eq:L-N} to relate the scale $L$ to the number of M2-branes as $L/\ell_{\text{P}} \sim N^{\frac{1}{6}}$. This shows that the logarithmic correction to the Euclidean path integral around the $-S^4\times S^7/\mathbb Z_k$ background is given by
\begin{equation}
    \Delta \log Z = \frac12\log N \, .\label{logZ:log}
\end{equation}
We emphasize that the coefficient of the $\log N$ term obtained in this way is independent of the orbifold order $k$ since the calculation above is valid for any choice of the $S^7/\mathbb{Z}_k$ internal space. This result matches the logarithmic correction obtained from the ABJM free energy~\eqref{eq:log-test}. Altogether,  this amounts to a nontrivial one-loop test of our general conjecture~\eqref{eq:micro-dS}.

\section{Discussion}\label{sec:discussion} \noindent
Based on explicit calculations, we have conjectured that the microscopic degrees of freedom behind the entropy of four-dimensional dS space can be encoded in the partition function of a 3d CFT on the round three-sphere that is dual to an auxiliary EAdS$_4$ saddle. Our reasoning exploits---and indeed advances---a chain of relations in semiclassical gravity that entwines $-S^4$ and EAdS$_4$ saddles as two real sections of a single complex geometry. We have shown that this effectively allows one to embed certain semiclassical de Sitter gravity theories in M-theory on $\mathrm{EAdS}_4 \times \mathrm{SE}^7$ backgrounds. In such embeddings, the dS entropy can be neatly expressed in terms of the regularized action of the EAdS domain. By AdS/CFT, the latter maps to the logarithm of the partition function of a holographic dual on the three-sphere which, we conjecture, encapsulates the exact dS entropy.

To test our conjecture, we have computed subleading and subsubleading quantum corrections to dS entropy induced by HD terms in the gravitational action and by one-loop effects. We have compared these with the expression for the ABJM partition function dual to the M-theory embedding on $S^7/\mathbb{Z}_k$. In effect, when the internal space is a smooth orbifold of the seven-sphere, the explicit form of the dual ABJM partition function as an Airy function yields detailed predictions for all higher-order corrections to the dS entropy, up to nonperturbative corrections. In the large $N$ expansion, the leading $N^{1/2}$ correction to the partition function corresponds to the effects of the HD terms on the gravity side. The next-order, $\log N$ logarithmic term in the partition function enters with a specific coefficient which, we have shown, matches with an independent one-loop calculation of this correction around the $-S^4 \times S^7/\mathbb{Z}_k$ background.

One may wonder whether one can also study the logarithmic correction to the dS entropy using a four-dimensional effective gravitational theory instead of the full 11d supergravity we employed here. This is a subtle question since the 11d $-S^4 \times S^7/\mathbb{Z}_k$ supergravity background is not scale separated which in turn means that the 4d effective theory is not a standard EFT coupled to gravity due to the presence of an infinite tower of Kaluza--Klein (KK) modes. This is in contrast to the approach taken in \cite{Anninos:2020hfj}, where similar logarithmic corrections were studied in the framework of general relativity coupled to a finite number of matter fields. It will be very interesting to understand how to reconcile these two alternative approaches to the calculation of the logarithmic corrections and in particular how to regularize the infinities arising from the presence of the tower of KK modes on $S^7/\mathbb{Z}_k$.

It is natural to consider other internal spaces different from $S^7/\mathbb{Z}_k$ to understand whether our conjecture extends beyond the ABJM theory. These different M-theory backgrounds will correspond to other 3d SCFTs that, based on the conjecture in \eqref{eq:micro-dS}, predict different corrections to the dS entropy. It is possible indeed to extend our ABJM analysis above to obtain explicit predictions for the dS entropy in such holographic models.

As a concrete example, consider a particular orbifold of the seven-sphere that is not freely acting but has fixed points. The internal space is denoted by $S^7/\mathbb{Z}_{N_f}$ and the corresponding 3d dual SCFT is known as the ADHM theory. This is an $\mathcal{N}=4$ U($N$) gauge theory with an adjoint hypermultiplet and $N_f$ fundamental hypermultiplets~\cite{Note5}.
The partition function of the ADHM theory on $S^3$ also takes a form similar to \eqref{eq:ABJMAiry} in terms of an Airy function~\cite{Mezei:2013gqa}. A holographic comparison between the $S^3$ free energy expanded to order $\mathcal{O}(N^{\frac{1}{2}})$ and the EAdS on-shell action, including four-derivative couplings, yields the holographic dictionary~\cite{Bobev:2021oku}
\begin{equation}
    \frac{L^2}{G_N} + 64\pi c_2 = \frac{2\sqrt{2N_f}}{3} N^{\frac{3}{2}} + \frac{N_f^2 - 4}{4\sqrt{2N_F}}N^{\frac{1}{2}} + o(N^{\frac{1}{2}}) \, .
\end{equation}
By the chain of dualities discussed in Section~\ref{sec:micro}, this determines the subleading corrections to the entropy of dS space---through its embedding in M-theory as an $-S^4 \times S^7/\mathbb{Z}_{N_f}$ background---in terms of the microscopic data $(N,N_f)$.

Another model that can be treated with similar methods corresponds to choosing as internal space  the $N^{0,1,0}/\mathbb{Z}_k$ manifold, which leads to an $\mathcal{N}=3$ 3d SCFT. Once again, the partition function of this SCFT on the three-sphere is controlled by an Airy function for a certain choice of parameters specifying the quiver~\cite{Marino:2011eh}. Using these results together with the partition function of the SCFT on another compact manifold, $S^1\times \Sigma_{\mathfrak{g}}$, one can find that the relevant holographic dictionary to order $\mathcal{O}(N^{\frac{1}{2}})$ reads, see \cite{Bobev:2023lkx},
\begin{equation}
    \frac{L^2}{G_N} + 64\pi\,c_2 = \frac{8\sqrt{k}}{3\sqrt{3}}\,N^{\frac{3}{2}} + \frac{k^2 - 4}{12\sqrt{3k}}\,N^{\frac{1}{2}} + o(N^{\frac{1}{2}}) \, .
\end{equation}
Using the line of reasoning in Section~\ref{sec:micro} this directly translates into a prediction for the quantum dS entropy associated with the $-S^4 \times N^{0,1,0}/\mathbb{Z}_k$ saddle of 11d supergravity, again valid to all orders in the large $N$ expansion.

Notably, the asymptotic expansion of the Airy function featuring in the ABJM, ADHM, and $N^{0,1,0}$ SCFTs discussed above leads to a universal $\frac14\log N$ term in the large $N$ expansion of the $S^3$ free energy. This can be combined with our conjecture to deduce that the logarithmic correction to the dS entropy of these M-theory models is also universal and given by \eqref{eq:log-test}. This nicely matches with our zero-mode counting in Section~\ref{sec:1loop}, which is insensitive to the details of the internal space in the eleven-dimensional embedding and thus in harmony with the universality of the logarithmic term. Given our conjecture \eqref{eq:micro-dS} and the nontrivial evidence for its validity presented above, it will be interesting to revisit large $N$ supersymmetric localization calculations on $S^3$ for 3d $\mathcal{N}=2$ SCFTs with a holographic dual description in M-theory in order to arrive at additional explicit examples for the microscopic dS entropy.

Another interesting avenue for future exploration is to extend our result in the context of the 3d ABJ holographic SCFT \cite{Aharony:2008gk}. This model admits a limit in which it connects with higher-spin gravity \cite{Chang:2012kt} and moreover allows for explicit calculations with the tools of supersymmetric localization, see for instance \cite{Binder:2020ckj}. It will be very interesting to understand whether our conjecture for the dS entropy can be extended to this model. Such a higher-spin dS/CFT would be distinct from the one discussed in \cite{Anninos:2011ui} and later in \cite{Hertog:2017ymy} which involves a nonunitary CFT. More broadly, the CFT duals in all examples discussed above are Euclidean counterparts of unitary theories, unlike the duals featuring in some models of dS/CFT based on the analytic continuation of theories in AdS/CFT. This highlights the crux of the route toward dS/CFT of \cite{Hertog:2011ky}, where complex saddles interlinking EdS and EAdS render obsolete the continuation from one classical background to another. The results in this paper serve not only as a one-loop test of our conjecture about the dS entropy, but also as an important off-shell test of that approach more generally.

Our proposal that holographically dual 3d CFT path integrals on the three-sphere encode the dS entropy yields a rather formal identification of the microscopic degrees of freedom that underpin this entropy. In particular, there is no obvious Lorentzian interpretation. Quite to the contrary, there is no obvious notion of time or even a Hamiltonian in the holographic dual on $S^3$, suggesting that it may be naive to think of the dS entropy as counting some microstates. This puzzling feature can be made even sharper if the conjecture in \eqref{eq:micro-dS} is taken at face value and pushed to the very quantum regime of finite $N$ and $k$. The ABJM partition function for low values of $N$ and $k$ can be computed explicitly and exactly, see \cite{Hatsuda:2012hm,Hatsuda:2012dt}. Analyzing these results in the context of the conjecture in \eqref{eq:micro-dS}, we are led to the conclusion that the exponential of the dS entropy is not an integer. This in turn strongly suggests that this entropy does not purely arise from counting some microscopic degrees of freedom but should rather be assigned a different interpretation. One may be tempted to speculate that this analysis points to the conclusion that the dS entropy is some kind of entanglement entropy, see \cite{Arias:2019pzy,Anninos:2020hfj} for a discussion of similar ideas. The presence of two holographically emergent copies of EAdS in the integration contour sketched in Figure~\ref{fig:contour} may also be viewed as circumstantial evidence for such an entanglement entropy interpretation. It will be very interesting to unpack these suggestive speculations further and make them more rigorous.

Finally, to get a better handle on dS space in Lorentzian signature, and deformations thereof, it would be very interesting to extend our analysis to more tangible cosmological observables. In this respect, it is worth noting that explicit expressions are available for the large $N$ partition function of the ABJM theory in the presence of various deformations that break conformal invariance, like squashings of the round $S^3$ or the addition of mass terms for scalar operators, see \cite{Bobev:2022jte,Hristov:2022lcw,Bobev:2022eus}. There are also corresponding semiclassical Euclidean supergravity saddles asymptotic to EAdS that are holographic duals to the deformed SCFT, see e.g.~\cite{Martelli:2011fu,Freedman:2013oja,Bobev:2018wbt}. These explicit results could presumably be used to add nontrivial sources in the gravitational path integral around a dS background in order to go beyond the zero-point function presented in this paper. This would pave the way to put earlier computations of cosmological observables in dS/CFT based on an ancillary AdS background, like the ones in \cite{Maldacena:2002vr,McFadden:2009fg,Hertog:2015nia}, on more solid ground. We plan to report on this elsewhere.

\section*{Acknowledgments} \noindent
We thank Dionysios Anninos, Anthony Charles, Kiril Hristov, and Thomas Van Riet for stimulating discussions. T.H.\ thanks Stephen Hawking and Jim Hartle for inspiring discussions about (quantum) cosmology on the sphere over many years. N.B.\ and J.H.\ are supported in part by Odysseus grant G0F9516N from the Research Foundation - Flanders (FWO). T.H.\ and J.K.\ are supported in part by the FWO research Grant G092617N. J.K.\ is also supported by doctoral fellowship 1171823N from the FWO. The work of V.R.\ is supported by a public grant as part of the Investissement d'avenir project, reference ANR-11-LABX-0056-LMH, LabEx LMH. We are also partially supported by the KU Leuven C1 grant ZKD1118 C16/16/005.

\vspace*{1mm}

\appendix

\section{Regularized on-shell higher-derivative action of EAdS\texorpdfstring{$_4$}{4}}
\label{App:regEAdS4} \noindent
In this appendix, we evaluate the regularized on-shell action of EAdS$_4$ in a gravitational theory with a negative cosmological constant and four-derivative couplings, by supplementing the action of the theory with the appropriate counterterms:
\begin{widetext}
\begin{align}
    \label{eq:EAdS-action}
    S =&\; \int d^4x\sqrt g \Bigl[-\frac{1}{16\pi G_N}\Bigl(R + \frac{6}{L^2}\Bigr) + (c_1 - c_2)C_{\mu\nu\rho\sigma}C^{\mu\nu\rho\sigma} + c_2\bigl(R_{\mu\nu\rho\sigma}R^{\mu\nu\rho\sigma}-4R_{\mu\nu}R^{\mu\nu}+R^2\bigr)\Bigr] \nonumber \\
    &\;+ \int d^3x \sqrt{h}\Bigl[\frac{1}{8\pi G_N}\Bigl(-K + \frac{2}{L} + \frac{L}{2}\mathcal{R}\Bigr) + (c_1 - c_2)\mathcal{L}^{\text{CT}}_{C^2} + 4c_2\bigl(\mathcal{J} - 2\mathcal{G}_{ab}K^{ab}\bigr)\Bigr] \, .
\end{align}
\end{widetext}
In the second line, we have included the counterterms needed for a well-defined variational principle and to regularize the divergences that arise in the large volume limit when one evaluates~\eqref{eq:EAdS-action} on-shell~\cite{Chamblin:1998pz,Myers:1987yn,Bobev:2021oku}. They are expressed in terms of the induced metric on the conformal boundary $h_{ab}$, the extrinsic curvature $K_{ab}$, and the Riemann tensor $\mathcal{R}_{abcd}$ of the induced metric. The quantity $\mathcal{J}$ is defined as the trace of the tensor
\begin{equation}
    3\mathcal{J}_{ab} = 2K K_{ac}K^c{}_b + K_{ab}K_{cd}K^{cd} - 2K_{ac}K^{cd}K_{db} - K^2K_{ab} \, ,\notag
\end{equation}
while $\mathcal{G}_{ab} = \mathcal{R}_{ab} - \frac12 h_{ab}\mathcal{R}$ denotes the boundary Einstein tensor. This theory admits an EAdS solution with line element
\begin{equation}
    \label{eq:EAdS4}
    ds^2 = dr^2+L^2\sinh^2(r/L)d\Omega_3^2 \, .
\end{equation}
Evaluating~\eqref{eq:EAdS-action} on the above solution gives rise to divergences due to the noncompactness of the space, which are regularized by the counterterms introduced in the second line. As shown in~\cite{Bobev:2021oku}, the divergences due to the Weyl-squared invariant can be regularized by adding a counterterm $\mathcal{L}^{\text{CT}}_{C^2}$ that is a linear combination of the Gibbons--Hawking and Gauss--Bonnet counterterms, but its explicit form will not be needed here since EAdS$_4$ is conformally flat.

Introducing a radial cut-off at $r=r_b$ and using the above definitions, we collect the various pieces contributing to the regularized on-shell action of EAdS$_4$. From the first line of~\eqref{eq:EAdS-action}, we find
\begin{equation}
    I_{1} = \Bigl(\frac{\pi L^2}{16G_N}+4\pi^2c_2\Bigr)\Bigl[8 - 9\cosh\Bigl(\frac{r_b}{L}\Bigr) + \cosh\Bigl(\frac{3r_b}{L}\Bigr)\Bigr] \, .
\end{equation}
From the counterterms on the second line, we find
\begin{align}
    I_{2}& = \frac{\pi L^2}{8 G_N}\sinh\Bigl(\frac{r_b}{L}\Bigr)\Bigl[4 + 2\cosh\Bigl(\frac{2r_b}{L}\Bigr) - 3\sinh\Bigl(\frac{2r_b}{L}\Bigr)\Bigr] \nn\\
    &\quad+ 4\pi^2 c_2\Bigl[9\cosh\Bigl(\frac{r_b}{L}\Bigr) - \cosh\Bigl(\frac{3r_b}{L}\Bigr)\Bigr] \, .
\end{align}
In total, the regularized on-shell action can be written as
\begin{equation}
    I_{\text{EAdS}} = \frac{\pi L^2}{8G_N}\Bigl(4 - e^{-3r_b/L} - 3\,e^{-r_b/L}\Bigr) + 32\pi^2 c_2 \, .
\end{equation}
Thus, we see that the counterterms precisely cancel the divergences that arise when the cut-off is sent to infinity, i.e.\ $r_b\rightarrow+\infty$, and we are left with
\begin{equation}
    \label{eq:I-AdS-E}
    I_{\text{EAdS}}^{\text{reg}} = \frac{\pi L^2}{2 G_N} + 32\pi^2 c_2 \, .
\end{equation}
We also note that the divergent pieces in the counterterm contribution $I_2$ are given by
\begin{equation}
    I_2^{\text{div}} = \Bigl(\frac{\pi L^2}{32G_N} + 2\pi^2 c_2\Bigr)e^{r_b/L}\bigl(9 - e^{2r_b/L}\bigr) \, ,
\end{equation}
which matches the divergent pieces in the real part of $I_\mathrm{L_{h}}$, given in~\eqref{eq:mI-4der-Lhor}, with $c_3 = 0$ as we have not considered the $R^2$ term in this appendix.

\bibliography{dSprx,footnotes}

\end{document}